\documentclass[aps,prl,10pt,twocolumn,showpacs,amsmath,amssymb,superscriptaddress]{revtex4-1}
\usepackage[english]{babel}
\usepackage[dvips]{graphicx}
\usepackage{dcolumn}
\usepackage{bm}
\usepackage{amsmath}
\usepackage{mathtools}
\usepackage{color}

\begin{document}
\title{Ultra-high brilliance multi-MeV $\gamma$-ray beam from non-linear Thomson scattering}
\author{G. Sarri}
\affiliation{School of Mathematics and Physics, The Queen's University of Belfast, BT7 1NN, Belfast, UK}
\author{D. J. Corvan}
\affiliation{School of Mathematics and Physics, The Queen's University of Belfast, BT7 1NN, Belfast, UK}
\author{W. Schumaker}
\affiliation{Center for Ultrafast Optical Science, University of Michigan, Ann Arbor, Michigan 48109-2099, USA}
\author{J. Cole}
\affiliation{The John Adams Institute for Accelerator Science, Imperial College of Science, Technology and Medicine, London SW7 2AZ, UK}
\author{A. Di Piazza}
\affiliation{Max-Planck-Institut f\"{u}r Kernphysik, Saupfercheckweg 1, 69117 Heidelberg, Germany}
\author{H. Ahmed}
\affiliation{School of Mathematics and Physics, The Queen's University of Belfast, BT7 1NN, Belfast, UK}
\author{C. Harvey}
\affiliation{School of Mathematics and Physics, The Queen's University of Belfast, BT7 1NN, Belfast, UK}
\author{C. H. Keitel}
\affiliation{Max-Planck-Institut f\"{u}r Kernphysik, Saupfercheckweg 1, 69117 Heidelberg, Germany}
\author{K. Krushelnick}
\affiliation{Center for Ultrafast Optical Science, University of Michigan, Ann Arbor, Michigan 48109-2099, USA}
\author{S. P. D. Mangles}
\affiliation{The John Adams Institute for Accelerator Science, Imperial College of Science, Technology and Medicine, London SW7 2AZ, UK}
\author{Z. Najmudin}
\affiliation{The John Adams Institute for Accelerator Science, Imperial College of Science, Technology and Medicine, London SW7 2AZ, UK}
\author{D. Symes}
\affiliation{Central Laser Facility, Rutherford Appleton Laboratory, Didcot, Oxfordshire OX11 0QX, United Kingdom}
\author{A. G. R. Thomas}
\affiliation{Center for Ultrafast Optical Science, University of Michigan, Ann Arbor, Michigan 48109-2099, USA}
\author{M. Yeung}
\affiliation{Helmholtz Institute Jena, Fr\"{o}belstieg 3, 07743 Jena, Germany}
\author {Z. Zhao}
\affiliation{Center for Ultrafast Optical Science, University of Michigan, Ann Arbor, Michigan 48109-2099, USA}
\author{M. Zepf}
\affiliation{School of Mathematics and Physics, The Queen's University of Belfast, BT7 1NN, Belfast, UK}
\affiliation{Helmholtz Institute Jena, Fr\"{o}belstieg 3, 07743 Jena, Germany}
\date{\today}
\begin{abstract}
We report on the generation of a narrow divergence ($\theta\approx 2.5$ mrad), multi-MeV ($E_\text{MAX} = 18$ MeV) and ultra-high brilliance ($\approx 2\times10^{19}$ photons s$^{-1}$ mm$^{-2}$ mrad $^{-2}$ 0.1\% BW) $\gamma$-ray beam from the scattering of an ultra-relativistic laser-wakefield accelerated electron beam in the field of a relativistically intense laser (dimensionless amplitude $a_0\approx2$). The spectrum of the generated $\gamma$-ray beam is measured, with MeV resolution, seamlessly from 6 MeV to 18 MeV, giving clear evidence of the onset of non-linear Thomson scattering. The photon source has the highest brilliance in the multi-MeV regime ever reported in the literature.
\end{abstract} 
\maketitle

The generation of high-quality Multi-MeV $\gamma$-ray beams is an active field of research due to the central role of these beams not only in fundamental research \cite{HIGS}, but also in extremely important practical applications, which include cancer radiotherapy \cite{cancerenergy,Gy}, active interrogation of materials \cite{interrogation}, and radiography of dense objects \cite{Glinec}. As an example, Giant Dipole Resonances of most heavy nuclei occur in an energy range of 15-30 MeV \cite{GDR}, exciting photo-fission of the nucleus. 

Different mechanisms have been proposed to generate $\gamma$-ray beams with these properties, including bremsstrahlung emission, synchrotron emission, and Compton scattering. Bremsstrahlung sources are routinely used for medical applications, and exploit electron beams accelerated by linear accelerators (LINAC) \cite{LINAC}. Laser-driven bremsstrahlung sources, whereby the electron beam is generated via laser-wakefield acceleration (LWFA) \cite{Esarey} have also been recently reported \cite{Glinec,Giulietti,Will}. However, the relatively broad divergence and source size limit the maximum brightness achievable with this technique and a more promising physical mechanism in this respect has been identified in Compton scattering. Laser-driven electron beams with energy per particle of the order of the GeV are now routinely available in the laboratory \cite{Esarey}, allowing for the possibility of all-optical and compact Compton-scattering sources \cite{Piazza,Corde}. 

Previous investigations of laser-driven Compton scattering have mostly focused on the linear regime, i.e. whenever the dimensionless intensity of the laser pulse is less than 1 ($a_0<$ 1, whereby $a_0 = eE_L/(m_e\omega_Lc)$, with $E_L$ and $\omega_L$ being the laser electric field and central frequency, respectively, and $m_e$ being the electron rest mass) \cite{Powers,Chen} and report on $\gamma$-ray energies ranging from a few hundreds of keV \cite{Powers} up to 3-4 MeV \cite{Chen}. Three main factors can in principle be modified in order to increase the energy of the generated photons: the electron Lorentz factor ($\gamma_e$), the laser photon energy $\hbar\omega_L$, or the laser intensity $a_0$.   The mean energy of the generated photons can in fact be estimated as:
\begin{equation}
E_\gamma \approx 4\gamma_e^2\hbar\omega_L f(a_0),\label{1}
\end{equation} 
with $f(a_0)\approx 1$ for $a_0\ll1$ and  $f(a_0)\approx a_0$ for $a_0\geq1$ \cite{Corde}.

\begin{figure}[!t]
\begin{center}
\includegraphics[width=1\columnwidth]{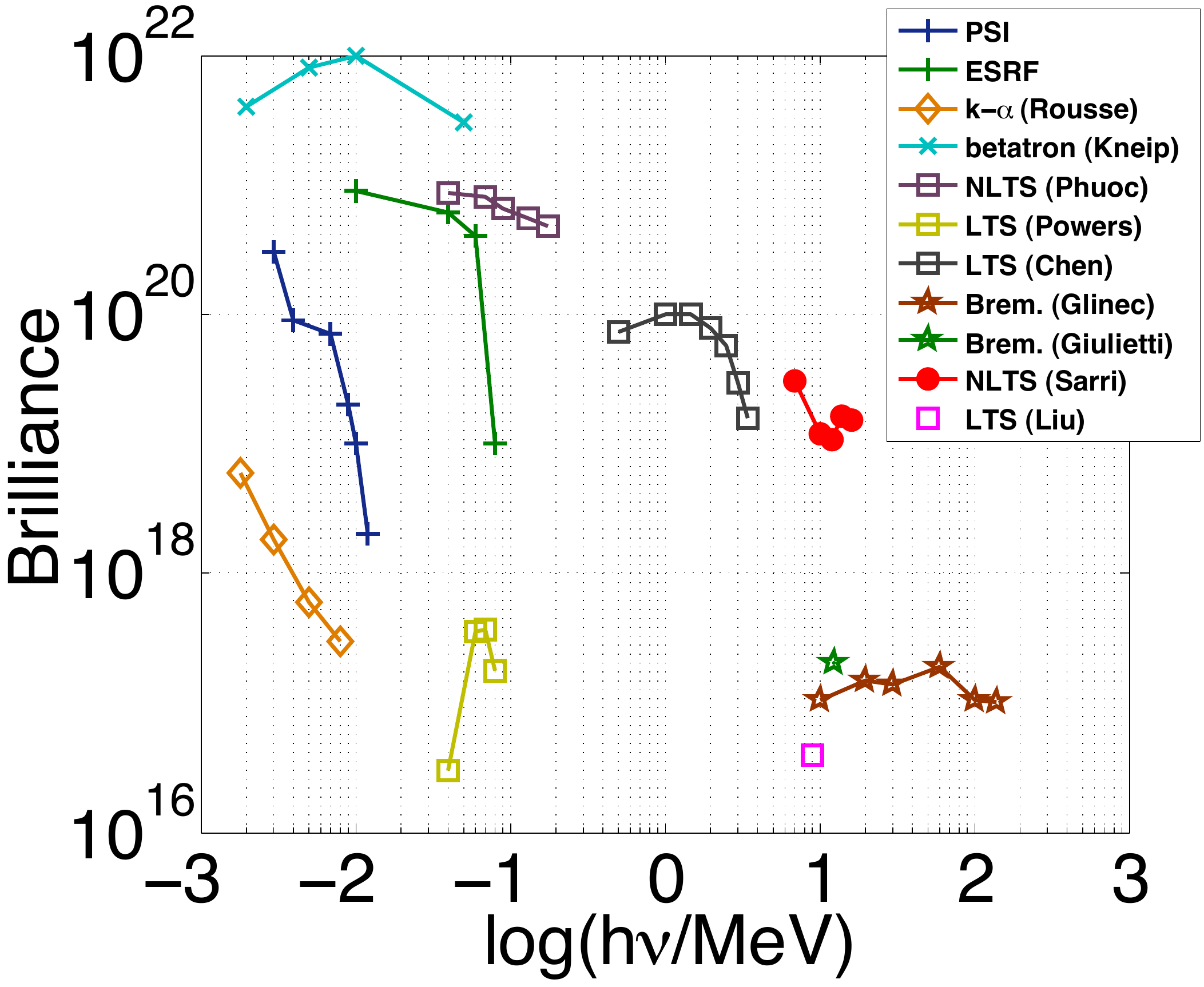}
\caption{Comparison of the present $\gamma$-ray source (solid red circles) with other generation mechanisms reported in the literature: k-$\alpha$ (orange diamonds, from Ref. \cite{Rousse}), solid-state undulators (green crosses from Ref. \cite{ESRF} and blue crosses from Ref. \cite{PSI}), betatron radiation (light blue crosses from Ref. \cite{Kneip}), bremsstrahlung radiation (green stars from Ref. \cite{Giulietti} and brown stars from Ref. \cite{Glinec}), and Thomson scattering (dark purple squares from Ref. \cite{Puhoc}, yellow squares from Ref. \cite{Powers}, grey squares from Ref. \cite{Chen}, and light purple square from Ref. \cite{Optlett}). Brilliance is expressed in units of photons s$^{-1}$ mm$^{-2}$ mrad $^{-2}$ 0.1\% BW} \label{comp}
\end{center}
\end{figure}

Liu and collaborators recently reported on an increase in photon energy (up to 8-9 MeV) by frequency converting the scattering laser up to its second harmonic (thus increasing $\hbar\omega_L$ in Eq. \ref{1}) \cite{Optlett}. However, using a higher laser frequency for scattering significantly reduces the laser energy available (crystal conversion efficiency into second harmonic of the order of 30 -50 \%), and the laser $a_0$ implying a modest number of generated photons ($\approx 3\times10^5$ photons per shot). This relatively low number can be easily understood if we consider that it would scale as $N_\gamma\propto a_0^2$ for $a_0\ll 1$ \cite{Corde}. The brilliance of this source is thus not higher than laser-driven bremsstrahlung source \cite{Glinec,Giulietti,Will} (see Fig. \ref{comp} for a comparison of reported brilliance for different $\gamma$-ray sources). Moreover, the $\gamma$-ray spectrum was not measured in this work but only inferred from numerical calculations. To the best of our knowledge, only one work has reported on non-linear laser-driven Compton scattering ($a_0\approx 1.5$) using a single laser to both drive and scatter the electrons \cite{Puhoc}. $\gamma$-ray energies of the order of few hundreds of keV were generated but the intrinsic difficulty in scaling this system to higher energies prevents it to be used for the generation of multi-MeV $\gamma$-ray beams. 

\begin{figure}[!t]
\begin{center}
\includegraphics[width=1\columnwidth]{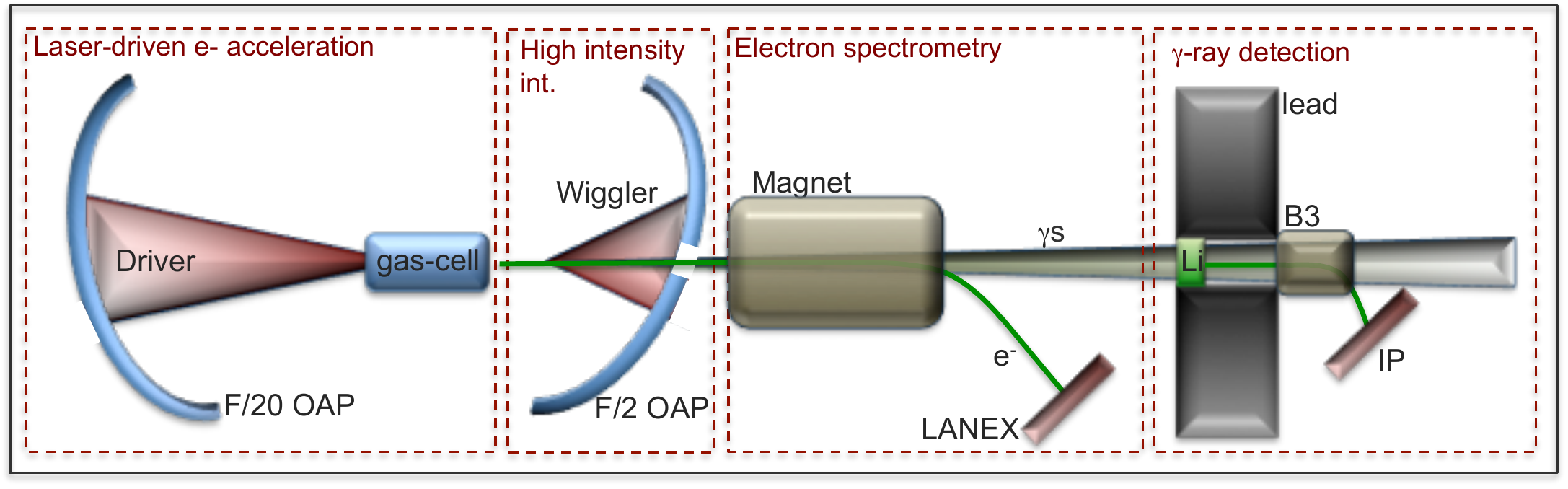}
\caption{Sketch of the experimental setup: a powerful laser pulse (\emph{Driver}) is focussed by an F/20 OAP at the edge of a gas-cell to generate an ultra-relativistic electron beam (first box: \emph{Laser-driven e- acceleration}). A second laser beam (\emph{Wiggler}) is focussed by a holed F/2 OAP counter-propagating to the electron beam (second box: \emph{High intensity int.}). After interaction, the electron beam is deflected by a strong pair of magnets onto a LANEX scintillator screen (third box: \emph{Electron spectrometry}) whilst the generated $\gamma$-ray beam propagates up to a Li-based spectrometer (fourth box: \emph{$\gamma$-ray  detection)}. An F/15 hole in the F/2 OAP ensures unperturbed propagation of the scattered electron beam and generated $\gamma$-ray beam  onto the detector, and minimises back-reflection of the laser in the amplification chain. } \label{setup}
\end{center}
\vspace{5mm}
\end{figure}

We report here on the generation of multi-MeV (maximum energy of the order of 16 - 18 MeV) and ultra-high brightness ($>10^7$ photons per shot with energy exceeding 6 MeV, implying a brightness exceeding $2\times10^{19}$ photons s$^{-1}$ mm$^{-2}$ mrad $^{-2}$ 0.1\% BW) following non-linear Compton scattering of ultra-relativistic laser-wakefield accelerated electron beam ($\gamma_e\approx 1000$) in the field of a ultra-intense laser pulse ($a_0\approx 2$, $\hbar\omega_L\approx 1.5$ eV). A novel $\gamma$-ray spectrometry technique allowed for the absolutely calibrated detection of the full spectrum of the $\gamma$-rays, clearly indicating onset of non-linear effects. To the best of our knowledge, this is the brightest $\gamma$-ray source in the multi-MeV energy range ever generated in a laboratory. 

\begin{figure}[!t]
\begin{center}
\includegraphics[width=1\columnwidth]{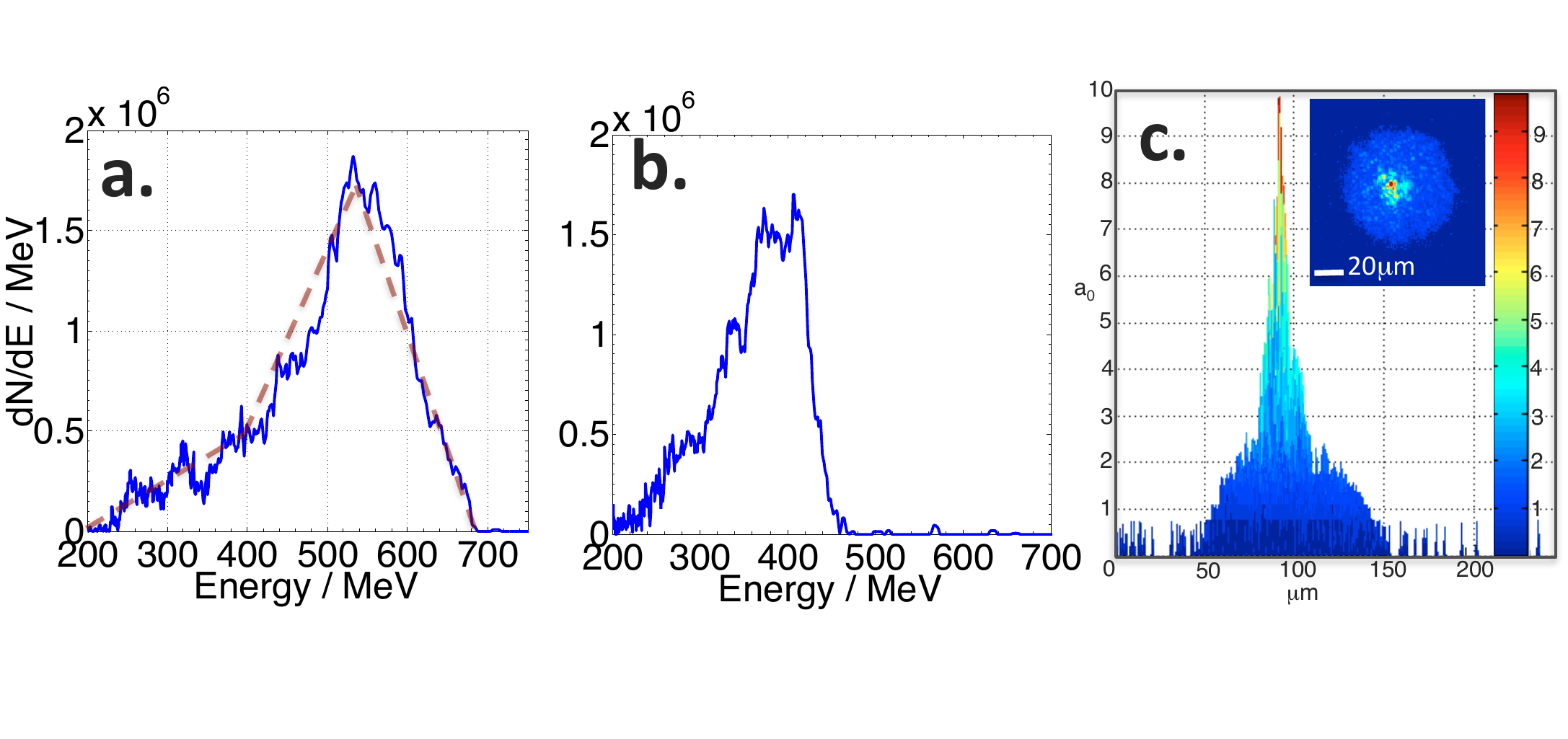}
\caption{\textbf{a.},\textbf{b.} Electron spectra used for the first and second series of experiments, respectively. In frame a., the dashed red line represents the approximated electron spectrum used as an input for the theoretical calculations. \textbf{c.} Measured intensity distribution of the laser used for scattering (\emph{Wiggler} in frame a. of Fig. \ref{setup}).} \label{electrons}
\end{center}
\vspace{5mm}
\end{figure}

The experiment was carried out using the Astra-Gemini laser, hosted by the Rutherford Appleton laboratory \cite{Hooker}, which delivers two laser beams each with central wavelength $\lambda_L\approx 800$ nm, pulse duration $\tau_L\approx (42\pm4)$ fs, and energy after compression of 18 J. Both lasers are generated from the same oscillator, avoiding problems of jitter in their synchronisation. One of the two laser beams was focussed, using an F/20 Off-Axis Parabola (OAP) down to a focal spot with Full Width Half Maximum of $27\pm3$ $\mu$m containing approximately 70\% of the laser energy (resulting intensity $I_\text{Driver}\approx 4\times10^{19} $ W/cm$^2$) at the edge of a 10mm long single-stage gas-cell filled with a mixture of He and N$_2$ (97\% - 3\%) at a pressure of 400 mbar. Once fully ionised, this pressure corresponds to an electron background density of $(3.2\pm0.2)\times10^{18}$ cm$^{-3}$ or, analogously, to a plasma period of the order of $\tau_\text{pl}\approx(60\pm2)$ fs, as measured via optical interferometry. This interaction produced, via laser-wakefield acceleration \cite{Esarey}, a quasi-monoenergetic electron beam with peak energy $E_{e^-}\approx$ 550 MeV (Lorentz factor $\gamma_{e^-}\approx 1.1 \times 10^3$) and a low energy tail, with a measured divergence of the order of 2 mrad ($16\pm3$ pC or $N_e=(1.0\pm0.2)\times10^8$ electrons in a $\pm 50$ MeV bandwidth around the electron peak energy, see Figs. \ref{electrons}.b and \ref{electrons}.c for typical electron spectra). For each run, the shot-to-shot fluctuation in electron beam energy and charge was consistently below 10\%. The shot-to-shot pointing fluctuation of the electrons was measured to occur with a standard deviation of the order of $0.7$ mrad \cite{Rory} and pulse front tilt effects on the electron beam axis (as first reported in Ref. \cite{Popp}) were carefully minimised prior to the experimental run \cite{Rory}. The second laser beam was instead focused, using an F/2 OAP with an F/15 hole in the middle, 1cm downstream of the exit of the gas-cell. At this point the electron beam diameter is measured to be  ($30\pm3$ $\mu$m). Insertion of a random spatial diffuser prior to the parabola allowed matching the electron beam transverse size with the high intensity focus of the laser with a peak dimensionless amplitude $a_0 =2$ (see Fig. \ref{electrons}.c). A peaked region with $a_0=10$ is also present (FWHM $\approx 3$ $\mu$m). however, due to its small spatial extent, only 1/100 of the electrons effectively interact with this higher intensity region. Numerical calculations (discussed in the following) indicate the contribution of this interaction to the $\gamma$ray spectrum to be negligible, and we will thus neglect it hereafter. The F/15 hole in the parabola was necessary in order to allow for clean transmission of the scattered electrons and the generated $\gamma$-ray beam and to avoid back-reflection of the two laser beams into the amplification chain. 

Downstream of the F/2 OAP, a pair of permanent magnets (B = 1 T, length of 15 cm) deflected the electrons away from the generated $\gamma$-ray beam to and a LANEX \cite{LANEX} scintillator. This arrangement allowed resolving electron energies from 120 MeV to 2 GeV. The LANEX scintillator was cross-calibrated using absolutely calibrated Imaging Plates \cite{IP}. An estimate of the quantum nonlinearity parameter $\chi_0=5.9\times 10^{-2}E_{e^-}[\text{GeV}]\sqrt{I_L[10^{20}\;\text{W/cm$^2$}]}$ \cite{Piazza} shows that in our experiment ($E_{e^-}\sim 500\;\text{MeV}$ and $I_L\sim8\times10^{18}\;\text{W/cm$^2$}$) quantum effects are negligible ($\chi_0\simeq$ 0.01). Moreover, by estimating the average energy $\mathcal{E}_{e^-}$ emitted by an electron with an energy of $500\;\text{MeV}$ from the Larmor formula \cite{Jackson}, we obtain $\mathcal{E}_{e^-}\sim 10\;\text{MeV}$ such that radiation-reaction effects are also negligibly small. The electron spectrum after the interaction is substantially unchanged  and it can thus be used as a valid approximation for the initial electron spectrum.  

Spatial overlap between the two laser pulses was achieved with 5$\mu$m precision using an alignment wire. Angular difference between the drive laser and the generated electron beam  \cite{Popp} ($\approx 1$ mrad) was measured using an electron beam profile imager and taken into account to achieve perfect laser-electron overlap \cite{Rory}. Temporal synchronisation was obtained using a spectral interferometry technique. Insertion of an 8\% reflection pellicle at the laser focal spots allowed sending both lasers through the F/15 hole of the F/2 OAP onto a diffraction grating. Measurement of the spectral fringes onto a CCD camera allowed synchronisation of the two lasers with a temporal resolution of the order of a few tens of fs. Details of the technique can be found in Ref. \cite{specint}.

The $\gamma$-rays were then spectrally resolved 2 meters downstream of the interaction. A 2cm thick block of Li (transverse size of 5 mm) was inserted in the $\gamma$-ray beam path to allow for the generation of secondary electrons via Compton scattering. This angular acceptance was explicitly chosen since it is comparable to the theoretically predicted angular spread of the $\gamma$-ray photons: $\theta_\gamma\approx a_0/\gamma_{e^-}\approx 2$ mrad. The on-axis scattered electron population retains the spectral shape of the $\gamma$-ray beam with an energy resolution of the order of the MeV. A 0.3 T, 5 cm long pairs of magnets spectrally dispersed the secondary electron beam onto an absolutely calibrated Imaging Plate \cite{IP}. This spectrometer was encased into a 30 cm thick box of lead to minimise noise arising from off-axis scattered electrons and photons and from bremsstrahlung photons emerging from the dumping of the primary electron beam. Typical spectral resolution, as resulting from the interplay of the magnetic spectrometer resolution and uncertainty introduced by the deconvolution process, was of the order of 10-15\% whereas the uncertainty in yield was of the order of 10\%. This system allowed us for the first time to measure the whole spectrum of the generated $\gamma$-ray beam, in an energy window between 6 and 20 MeV and an energy resolution of the order of 1 MeV. A detailed description of this spectrometer, including the procedure adopted for the deconvolution of the spectrum of the secondary electron beam, can be found in Ref. \cite{Corvan}.

\begin{figure}[!h]
\begin{center}
\includegraphics[width=0.8\columnwidth]{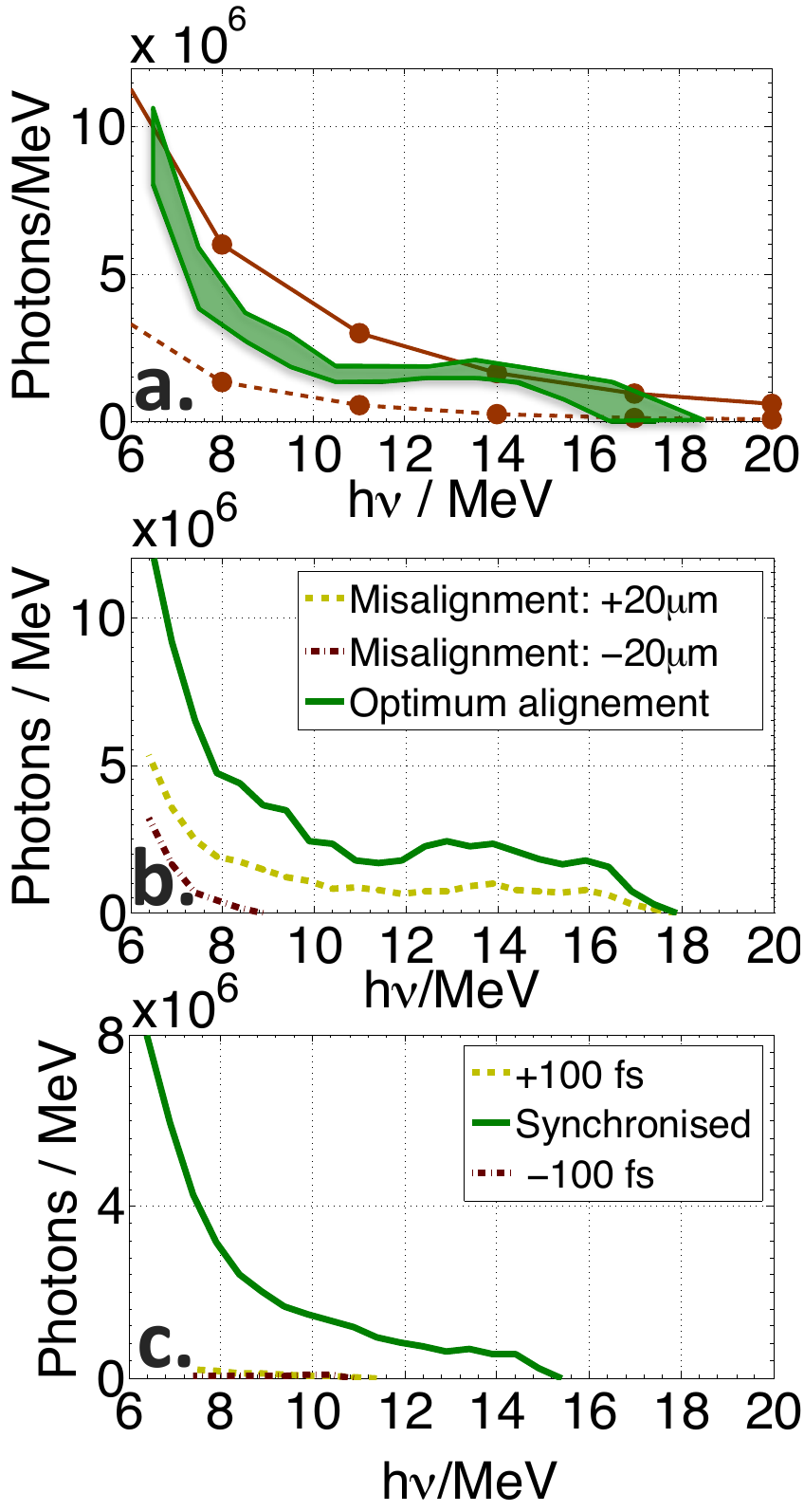}
\caption{\textbf{a.} green band: $\gamma$-ray spectrum as measured during the interaction of the laser-driven electron beam (spectrum depicted in Fig. \ref{electrons}.a) with the high-intensity focal spot of a secondary laser beam (spatial distribution shown in Fig. \ref{electrons}.c). The band represents the uncertainty associated with the experiment, as mainly resulting from the spectral resolution of the $\gamma$-ray spectrometer and the response of the detector. Solid and dashed brown lines depict theoretical expectations for the same electron and laser parameter as the experimental ones but with $a_0=2$ and $a_0=1$, respectively. \textbf{b.} Effect of the spatial misalignment on the $\gamma$-ray yield. The green line represents the measured spectrum for optimised electron-laser overlap (same as green line in frame a.) whereas dashed curves depicts the measured spectra if an artificial misalignment of $\pm$20 $\mu$m is introduced. \textbf{c.} Effect of the temporal synchronisation on the $\gamma$-ray yield. The green line represents the measured spectrum for optimised electron-laser synchronisation (for an electron spectrum as the one in Fig. \ref{electrons}.b) whereas dashed curves depict the measured spectra if an artificial desynchronisation of $\pm$100 fs is introduced.} \label{spectra}
\end{center}
\end{figure}

In this paper, we will discuss the experimental results obtained in two different runs. In both runs a consistent electron beam (shot-to-shot fluctuation in beam energy and charge of less than 10\% in both cases) was obtained, with a typical spectrum as the one depicted in Fig. \ref{electrons}.a for the first run and with a typical spectrum as the one depicted in Fig. \ref{electrons}.b for the second run. In conditions of best overlap and synchronisation between the electron beam and the laser pulse, the $\gamma$-ray spectrum exhibits a monotonically decreasing profile, with a typical number of photons per MeV of the order of $10^6$, extending up to 15-18 MeV (green band in Fig. \ref{spectra}.a obtained with the electron spectrum shown in Fig. \ref{electrons}.a). To ensure that the recorded signal arises from the electrons wiggling in the laser beam focus, we varied their temporal and spatial overlap. An artificially induced spatial misalignment of $\pm 20 \mu$m significantly reduces the signal whilst roughly preserving the same spectral shape (Fig. \ref{spectra}.b obtained with the electron spectrum shown in Fig. \ref{electrons}.a). No signal can be recorded if the misalignment is further increased to $\pm 40 \mu$m. Also, changing the relative delay of the lasers of $\pm 100$fs reduces the signal virtually to zero (Fig. \ref{spectra}.c obtained with the electron spectrum shown in Fig. \ref{electrons}.b). The number and peak energy of the measured $\gamma$-ray photons is in good agreement with synchrotron calculations for $a_0=2$ (see Fig. \ref{spectra}.a). 

The theoretical emission spectra have been obtained by numerically integrating the classical Lorentz equation in the presence of a plane wave laser field, with a  Gaussian temporal profile and an initial electron spectrum as the one depicted by the red dashed line in Fig. \ref{setup}.b. Indeed, the parameters characterising the laser field and the electron beam in the experiment are such that the plane wave approximation works reasonably well and that both quantum and radiation-reaction effects can be neglected \cite{Piazza}. Once the electron trajectories are determined, they have been used to calculate the radiated electromagnetic fields via the Li\'{e}nard-Wiechert potential \cite{Jackson}. Finally, the obtained angular-resolved energy spectra have been integrated with respect to the emission angles according to the experimental condition (half-cone angle from 0 to $1.25\;\text{mrad}$).  This procedure has been carried out for $a_0=1$ and $a_0=2$ (dashed and solid brown lines in Fig. \ref{spectra}.a). It is interesting to note that in a linear regime ($a_0=1$, see Fig. \ref{spectra}.a) the same setup would ensure a much lower peak energy and number of photons, clearly confirming the better performance of non-linear scattering for the generation of high brilliance and high energy $\gamma$-ray beams. 

Let us now proceed to estimate the peak brilliance of our source. The source size is comparable to the electron beam diameter at interaction ($D_\gamma$ $\approx30\pm3\mu$m), whereas un upper limit for divergence of the measured photon beam is given by the angular acceptance of the $\gamma$-ray spectrometer (2.5 mrad). Finally, the temporal duration would be comparable to that of the electron beam and, therefore, of the order of half plasma period in the acceleration stage \cite{Mangles} ($\approx$ 30 fs for a plasma period of $\tau_\text{pl}\approx (60\pm2)$ fs). In a 0.1\% bandwidth around 15 MeV we have approximately $(3.0\pm0.2)\times10^3$ photons, implying a lower limit for the peak brilliance of $(1.8\pm0.4)\times10^{19}$ photons s$^{-1}$ mm$^{-2}$ mrad$^{-2}$ 0.1\% BW. This brilliance is the highest ever achieved in a laboratory for multi-MeV $\gamma$-ray sources exceeding by approximately two orders of magnitude that achieved by bremsstrahlung sources (see Fig. \ref{comp}). This is due to the unique combination of high photon energy (maximum of 15 - 18 MeV compared to a sub-MeV for betatron and a few MeV for linear Thomson scattering sources), high photon number (approximately $10^7$ photons with energy exceeding 6 MeV per laser shot), small divergence and source size (2.5 mrad and 30 $\mu$m compared to tens of mrad and hundreds of microns for typical bremsstrahlung sources, respectively), and short duration (tens of fs, compared to picoseconds or nanoseconds for solid-state systems). 

In conclusion, we report on experimental evidence of non-linear Thomson scattering in a two-laser counter-propagating geometry. The absolutely calibrated spectrum of the generated $\gamma$-ray beam has been seamlessly measured, with MeV resolution, from 6 to 20 MeV and provides clear experimental evidence of non-linear Thomson scattering. Thanks to the high photon number generated, short duration, narrow divergence, and small source size, this photon source presents the highest brilliance ever obtained in the laboratory in a multi-MeV energy window.  

\textbf{Acknowledgments}\\
G.S. and M.Z. wish to acknowledge the EPSRC grants EP/L013975/1 and EP/I029206/1. 


\end{document}